\documentstyle[12pt]{article}
\newcommand{\beq}{\begin{equation}}
\newcommand{\eeq}{\end{equation}}
\begin{document}
\title
{\bf The Electromagnetic Effects In Isospin Symmetry Breakings 
Of $q\bar q$ Systems}

\author{Shi-lin Zhu, and Zhenping Li  \\
      Department of Physics, Peking University \\
 Beijing 100871, P.R.China}

\maketitle
\indent
\begin{abstract}
\baselineskip=24pt
The isospin symmetry breakings of $q\bar q$ are investigated in the 
QCD sum rule method.  The electromagnetic effects are evaluated following 
the procedure requiring that the electromagnetic effects for charged meson
be gauge invariant.  We find that  
the electromagnetic effects are also dominant in the isospin violations
of the $\rho$ mesons, which have been shown to be 
the case in the mass splittings of pions. The numerical results for the
differences of pion decay constants and the masses of $\rho$ mesons
are presented, which are consistent with the data.
\end{abstract}

\noindent PACS number(s): 13.40.Dk, 11.30.Hv, 12.38.Lg
 
\newpage
\baselineskip=24pt
The origin of mass differences in isospin multiplets has long been of great
interest in nuclear and particle physics as a source of information about
symmetry violations.  
Hadronic isospin violations are particularly important in that they arise from
quark mass differences in Quantum Chromodynamics (QCD) 
(see Ref.\cite{gl} for a review of the early work in this area), 
and of course 
electromagnetic effects.  Among the first applications of the method of QCD 
sum rules was the
study of isospin violations in the $\rho-\omega$ system\cite{svz}, in which it
was recognized that the isospin splitting of the light-quark condensates can
produce effects as large as the current-quark mass splittings and
electromagnetic effects.  
Recently the QCD sum rule method has been used to 
study the neutron-proton mass difference\cite{hhp,yhhk}, the octet baryon
mass splittings\cite{ada} and the mass differences in the charmed meson
systems (the D and $\rm{D}^*$ scalar and vector mesons)\cite{ei}.  

The isospin symmetry breakings of pions have been discussed extensively in the
chiral perturbation theory, in which the electromagnetic effects are shown to 
be dominant in the mass splittings of pions.  Thus,  a consistent treatment
of the electromagnetic effect
 is particularly important for the isospin 
symmetry breakings in the $q \bar q$, $q=u$ or $d$, systems.  
In comparison with hadronic quark models, the QCD sum rule method for 
calculating isospin splittings has the advantage that one can directly
use QED field theory rather than rely on models to estimate Coulomb
corrections.  In Ref. \cite{gauge}, a systematic approach for the 
electromagnetic effects in the operator product expansion for
 the $q\bar q$ systems has been developed, in which the problem of the gauge
invariance for the charged $q\bar q$ states was resolved.  This enables us
to investigate the violations of the isospin symmetry due to all three effects,
the nonperturbative effects, the quark mass differences and the electromagnetic
effects, consistently in the QCD sum rule method. 
The nonterturbative effect is due to the difference of the quark condensate,  
which arises from the up and down quark mass difference.
The results of our 
studies\cite{hls} in the heavy-light quark systems are in good agreement 
with the available data.  The focus of this paper is to extend our early 
studies to the light-light quark systems, such as the $\pi$ and $\rho$ 
systems.

The basic approach of the QCD sum rule is to study the two-point correlation function in the Wilson operator product expansion (OPE), defined by
\begin{eqnarray}\label{i}
\bar \Pi_{\mu\nu}(q^2) & = & i\int d^4x e^{iqx} <T(J_{\mu}(x)\bar
J_{\nu}(0))> 
\nonumber\\
 & = & \Pi (q^2) (q_\mu q_\nu -q^2 g_{\mu\nu}) 
  +\Pi_1(q^2) q_\mu q_{\nu} \, \, ,
\end{eqnarray}
%\begin{eqnarray}\label{i1}
%\bar \Pi^5_{\mu\nu}(q^2) & = & i\int d^4x e^{iqx} <T(J^5_{\mu}(x)\bar
%J^5_{\nu}(0))> 
%\nonumber\\
% & = & \Pi^5 (q^2) q_\mu q_\nu + \Pi^5_1(q^2) q_\mu q_{\nu} 
%\end{eqnarray}
where 
\begin{equation}\label{ii}
J_{\mu}(x) = {\bar q_1}(x)\gamma_\mu q_2(x).
\end{equation}
%and for the pseudo-vector current 
%\begin{equation}\label{iii}
%J^5_{\mu}(x) = {\bar q_1}(x)\gamma_\mu\gamma_5 q_2(x).
%\end{equation}
%%%%%%%%%%%

The $\Pi (q^2)$ in Eq. \ref{i} can be written as
\begin{equation}\label{1}
\Pi (q^2) =\Pi_0 (q^2) +\Pi_{\mbox{em}} (q^2) \  ,
\end{equation}                  
where $\Pi_0 (q^2)$ is the leading term, and $\Pi_{\mbox{em}}$ represents the
contributions from the electromagnetic effects.  The quantity  $\Pi_0 (q^2)$
has been evaluated in the literature. Thus we extend  
the results of $\Pi_0(q^2)$ in Refs. \cite{svz,reinders} by including the flavor dependences of each term. 
In order to study the violations of the isospin symmetry,
the electromagnetic effects $\Pi_{\mbox{em}} (q^2)$ should also be written 
in the framework of the operator product expansion.  
The leading electromagnetic effects in this approach are the
two-point functions from a two-loop perturbative contribution, whose 
Feynman diagrams are shown in Fig. 1.  For the charge neutral 
current $\Pi_{\mbox{em}}^0 (q^2)$, one could simply obtain  the two-point functions by changing the gluons in QCD to the photons
in QED in Fig. 1, since the two-point functions has been calculated 
in QCD\cite{svz,reinders}. 
\begin{equation}
\Pi_{\mbox{em}}^0 (q^2) = 
-{1\over 4\pi^2} e_q^2 {\alpha_e \over \pi} 
\ln {-q^2 \over \mu^2} \, ,
\end{equation}
where $e_q$ is the quark charge and $\mu$ is the infrared cutoff.
Obviously, this result is gauge invariant.

Following the procedure in Ref. \cite{gauge}, there are additional 
Feynman diagrams shown in Fig.2 for the charged current so that the 
$\Pi_{\mbox{em}}^{\pm} (q^2)$ for the charged $q\bar q$ systems is gauge
invariant.  Thus, the two point function $\Pi_{\mbox{em}}^{\pm} (q^2)$ for 
the charged $q\bar q$ systems is 
\begin{equation}
\Pi_{\mbox{em}}^{\pm}  (q^2)  =
-{1\over 4\pi^2} e_{q_1} e_{q_2} {\alpha \over \pi} 
\ln {-q^2 \over \mu^2} \, .
\end{equation}

The leading nonperturbative terms due to the electromagnetic effects are 
the four quark condensates, of which the Feynman diagrams are shown in 
Fig. 3 and Fig. 4. We present the expressions for $\Pi (q^2)$ 
\begin{eqnarray}
\label{rho-}
\Pi_{\rho^{\pm}}  = 
- {1\over 4\pi^2} (1+{\alpha_s \over \pi}) \ln {-q^2 \over \mu^2}
- {1\over 4\pi^2} e_u e_d{\alpha \over \pi} \ln {-q^2 \over \mu^2} \nonumber
\\
+{1\over q^4} (m_u \langle {\bar d}d\rangle + m_d \langle {\bar u}u\rangle )
+{1\over 12q^4}\langle {\alpha_s \over \pi} G^2 \rangle \nonumber \\
+{8 \over 9 q^6}  g_s^2 \langle {\bar u} u \rangle 
\langle {\bar d} d \rangle 
-{8 \over 81 q^6} g_s^2 (\langle  {\bar u} u \rangle^2 
+ \langle {\bar d} d \rangle^2 )  \nonumber \\
+{7 \over 27 q^6}e^2 e_u e_d (\langle {\bar u} u \rangle^2 
+\langle {\bar d} d \rangle^2 ) 
+{1\over 4q^6}e^2 e^2_T \langle {\bar u} u \rangle 
\langle {\bar d} d \rangle 
,
\end{eqnarray}
and
\begin{eqnarray}
\label{rho-0}
\Pi_{\rho^0} = 
- {1\over 4\pi^2} (1+{\alpha_s \over \pi} )\ln {-q^2 \over \mu^2}
- {1\over 4\pi^2} {e_u^2 + e_d^2 \over 2}{\alpha \over \pi} \ln {-q^2 \over \mu^2} \nonumber \\
+{1\over q^4} (m_d \langle {\bar d}d\rangle + m_u \langle {\bar u}u\rangle )
+{1\over 12q^4}\langle {\alpha_s \over \pi} G^2 \rangle \nonumber \\
+{28 \over 81 q^6} g_s^2 (\langle {\bar u} u \rangle^2 
+\langle  {\bar d} d \rangle^2 ) 
+{7 \over 27 q^6} e^2 (e_u^2 \langle  {\bar u} u \rangle^2 
+e_d^2 \langle {\bar d} d \rangle^2 ) 
 ,
\end{eqnarray}
for the $\rho$ systems, and 
\begin{eqnarray}
\label{pi-}
\Pi_{\pi^{\pm}} = 
- {1\over 4\pi^2} (1+{\alpha_s \over \pi}) \ln {-q^2 \over \mu^2}
- {1\over 4\pi^2} e_u e_d{\alpha \over \pi} \ln {-q^2 \over \mu^2}\nonumber
 \\
+{1\over q^4} (m_u \langle {\bar d}d\rangle + m_d \langle {\bar u}u\rangle )
+{1\over 12q^4}\langle {\alpha_s \over \pi} G^2 \rangle \nonumber \\
-{88 \over 81 q^6}  g_s^2 \langle {\bar u} u \rangle 
\langle {\bar d} d \rangle 
-{11 \over 27 q^6} e^2 e_u e_d (\langle  {\bar u} u \rangle^2 
+ \langle {\bar d} d \rangle^2 )    ,
\end{eqnarray}
and 
\begin{eqnarray}
\label{pi-0}
\Pi_{\pi^0}  = 
- {1\over 4\pi^2} (1+{\alpha_s \over \pi}) \ln {-q^2 \over \mu^2}
- {1\over 4\pi^2} {e_u^2 + e_d^2 \over 2}{\alpha \over \pi} \ln {-q^2 \over \mu^2} \nonumber \\
+{1\over q^4} (m_d \langle {\bar d}d\rangle + m_u \langle {\bar u}u\rangle )
+{1\over 12q^4}\langle {\alpha_s \over \pi} G^2 \rangle \nonumber \\
-{44 \over 81 q^6} g_s^2 (\langle  {\bar u} u \rangle^2 
+\langle  {\bar d} d \rangle^2 ) 
-{11 \over 27 q^6} e^2 (e_u^2 \langle  {\bar u} u \rangle^2 
+e_d^2 \langle {\bar d} d \rangle^2 ) 
\end{eqnarray}
for the $\pi$ systems, in which  we have assumed the vacuum dominance 
and factorization hypothesis for the four quark condensates 
following Ref. \cite{svz}. In the Appendix we present expressions in which 
the vacuum dominance approximation has not been implemented. The sensitivity 
to the possible deviation from vacuum dominance approximation will be 
discussed later.

The isospin symmetry breakings of these systems are determined by the 
difference between $\Pi^{\pm}(q^2)-\Pi^0(q^2)$, and we have 
\begin{eqnarray}\label{d1}
\Pi_{\rho^0} -\Pi_{\rho^\pm} =  
-{1\over 8\pi^2} (e_u -e_d)^2 {\alpha \over \pi} \ln {-q^2\over \mu^2} 
+ {1\over 27q^6} \pi\alpha (e_u -e_d)^2 
\langle {\bar u} u \rangle^2 \nonumber \\
+{4\over 9q^6} g_s^2 
(\langle {\bar u} u \rangle -\langle {\bar d} d \rangle )^2
 +{1\over q^4} (m_u -m_d) 
(\langle {\bar u} u \rangle -\langle {\bar d} d 
\rangle ) 
\end{eqnarray}
for $\rho$ states, and   
\begin{eqnarray}\label{d2}
\Pi_{\pi^0} -\Pi_{\pi^\pm}=  
-{1\over 8\pi^2} (e_u -e_d)^2 {\alpha \over \pi} \ln {-q^2\over \mu^2} 
- {44\over 27q^6} \pi\alpha (e_u -e_d)^2 
\langle {\bar u} u \rangle^2  \nonumber \\
 -{44\over 81q^6} g_s^2 
(\langle {\bar u} u \rangle -\langle {\bar d} d \rangle )^2
 +{1\over q^4} (m_u -m_d) 
(\langle {\bar u} u \rangle -\langle {\bar d} d 
\rangle ) 
\end{eqnarray}
for pions.  The first two terms in Eqs. \ref{d1} and \ref{d2} represents
the electromagnetic effects, and the last two terms correspond to the isospin
violations in the quark masses and condensates.  Qualitatively, the 
contributions from the electromagnetic effects and the differences of the quark
masses and condensates to the isospin symmetry breakings of $\pi$ and $\rho$
states should have the same order of magnitudes.  Thus, Eqs. \ref{d1} and \ref{d2} show explicitly that the contributions from the differences of 
quark masses and condensates are of higher order relative to the 
electomagnetic effects.  This has been shown in Ref. \cite{gl} for pions 
sometime ago,  while Eq. \ref{d1} suggests that the electromagnetic 
effects is also dominant in the isospin symmetry breakings of $\rho$ mesons.  

We adopt the standard one resonance plus a continuum 
model for the spectral density Im$\Pi_{\rho^0} (s)$ at the hadronic 
level:
\begin{equation}
\mbox{Im} \Pi_{\rho^0} (s) = \pi f^2_{\rho^0} \delta (s -m^2_{\rho}) + 
{1\over 4\pi} (1+ {\alpha_s \over \pi} + 
{e_u^2 +e_d^2 \over 2} {\alpha \over \pi} )\theta (s- s_{\rho^0})
\end{equation}
where $f_{\rho^0}$ is related to the electronic width of the $\rho$ meson and 
$s_{\rho^0} \approx 1.5$GeV$^2$ is the $\rho$ meson continuum threshold. 
We use $s_{\pi} \approx 0.75$GeV$^2$ in the pseudo vector channel so that
the sum rule is dominated by the pion and 
the $a_1(1260)$ meson contributes only to the continuum. Although it is 
impossible to extract the pion masses in the QCD sum rule framework, 
it is still possible to calculate their decay constants with the physical
 masses as the inputs \cite{svz,reinders}. 
After making the Borel transformation           
and transferring the continuum contribution to the right hand side,  
 the final sum rules are:
\begin{eqnarray}
\label{r-}
f^2_{\rho^{\pm}} e^{ -{m^2_{\rho^{\pm}} \over M_B^2} }
 =  {1\over 4\pi^2} \{ 
(1+{\alpha_s(M_B^2) \over \pi} + e_u e_d{\alpha \over \pi} ) 
M_B^2 (1- e^{- {s_{\rho^{\pm}}\over M_B^2}}) \nonumber \\
-{ m_u a_d + m_d a_u \over M_B^2} 
+{\langle {\alpha_s \over \pi} G^2 \rangle  \over 12 M_B^2} 
-{4\over 9} {{\bar \alpha_s} \over \pi} {a_u a_d \over M_B^4} \nonumber \\
+{4\over 81} {{\bar \alpha_s} \over \pi} {a_u^2 + a^2_d \over M_B^4}
-{7\over 54} e_u e_d {\alpha \over \pi} {a_u^2 + a^2_d \over M_B^4}
-{1\over 8} e_T^2 {\alpha \over \pi} {a_u a_d \over M_B^4} \},
\end{eqnarray}
and
\begin{eqnarray}
\label{r0}
f^2_{\rho^0} e^{ -{m^2_{\rho^0} \over M_B^2} }
 =  {1\over 4\pi^2} \{ 
(1+{\alpha_s(M_B^2) \over \pi} + {e^2_u +e^2_d \over 2}{\alpha \over \pi} ) 
M_B^2 (1- e^{- {s_{\rho^0}\over M_B^2}})
\nonumber \\
-{ m_u a_u + m_u a_u \over M_B^2} 
+{\langle {\alpha_s \over \pi} G^2 \rangle  \over 12 M_B^2} \nonumber \\
-{14\over 81} {{\bar \alpha_s} \over \pi} {a_u^2 + a^2_d \over M_B^4}
-{7\over 54} {\alpha \over \pi} {e_u^2 a_u^2 + e_d^2 a^2_d \over M_B^4} \}
\end{eqnarray}
for $\rho$ mesons, and
\begin{eqnarray}
\label{p-}
f^2_{\pi^{\pm}} e^{ -{m^2_{\pi^{\pm}} \over M_B^2} }
=  {1\over 4\pi^2} \{ 
(1+{\alpha_s(M_B^2) \over \pi} + e_u e_d{\alpha \over \pi} ) 
M_B^2 (1- e^{- {s_{\pi^{\pm}}\over M_B^2}}) \nonumber \\
-{ m_u a_d + m_d a_u \over M_B^2} 
+{\langle {\alpha_s \over \pi} G^2 \rangle  \over 12 M_B^2} \nonumber
 \\
+{44\over 81} {{\bar \alpha_s} \over \pi} {a_u a_d \over M_B^4} 
+{11\over 54} e_u e_d {\alpha \over \pi} {a_u^2 + a^2_d \over M_B^4} ,
\end{eqnarray}
and 
\begin{eqnarray}
\label{p0}
f^2_{\pi^0} e^{ -{m^2_{\pi^0} \over M_B^2} }
 =  {1\over 4\pi^2} \{ 
(1+{\alpha_s(M_B^2) \over \pi} + {e^2_u +e^2_d \over 2}{\alpha \over \pi} ) 
M_B^2 (1- e^{- {s_{\pi^0}\over M_B^2}}) \nonumber \\
-{ m_u a_u + m_u a_u \over M_B^2} 
+{\langle {\alpha_s \over \pi} G^2 \rangle  \over 12 M_B^2} \nonumber \\
+{22\over 81} {{\bar \alpha_s} \over \pi} {a_u^2 + a^2_d \over M_B^4}
+{11\over 54} {\alpha \over \pi} {e_u^2 a_u^2 + e_d^2 a^2_d \over M_B^4} \}
\end{eqnarray}
where $e_T =e_u -e_d$, 
$e_u ={2\over 3}$, $e_d  =-{1\over 3}$, 
$m_u =5.1$MeV, $m_d =8.9$MeV,  
$a_q =-(2\pi )^2 \langle {\bar q} q \rangle$, q=u, d,  
$\alpha ={1 \over 137}$, 
$\alpha_s (M_B^2) = {4\pi \over \ln (100M_B^2)} $, ${\bar \alpha_s} =0.7$,
and the values of the various condensates $a_q$ are standard\cite{svz} 
 ${\bar a} ={a_u +a_d \over 2} =0.55$GeV$^3$, 
and $\langle g_s^2 G^2 \rangle =0.474$GeV$^4$.  

The mass sum rules for $\rho$ mesons are obtained by taking the ratio of 
Eq. (\ref{r-}) and the resulting equation of the first 
derivative with respect to ${1 \over M_B^2}$.  The  
$\Delta m_{\rho} $ is shown in Fig. 5.  The threshold parameters for the
stable sum rules are $s_{\rho^{\pm}} =1.502$GeV$^2$ and $s_{\rho^0} 
=1.500$GeV$^2$.  The mass of
$\rho^{\pm}$ is found to be $770$ MeV, which reproduces the results in Refs. 
\cite{svz,reinders} and is in good agreement with data.
Our numerical analysis shows that contributions from the differences 
of quark masses and condensates are indeed negligible.  We have
the resulting $\Delta m_{\rho} =-0.25$ MeV,  and this is not 
inconsistent with the experimental value from the Particle Data Group,  
of which  $\Delta m_{\rho} =0.3\pm 2.2$MeV\cite{pdg}. 

With the physical pion masses $m_{\pi^{\pm}}=139.57$MeV and 
$m_{\pi^0}=134.98$MeV as inputs,  one could extract the decay 
constants of pions.  The decay constant $f_{\pi^{\pm}}$ from these
sum rules is $131$ MeV, and we present the 
sum rule for $\Delta f_{\pi} =f_{\pi^{\pm}}-f_{\pi^0}$ in Fig. 6. 
The corresponding threshold parameters are  $s_{\pi^{\pm}}=0.78$GeV$^2$ 
and $s_{\pi^0}=0.70$GeV$^2$.  Our result is  $\Delta f_{\pi} =4.5$MeV, 
while the data in the Particle Data Group for $f_{\pi}$ suggest 
$\Delta f_{\pi} =11.9\pm 4.0$ MeV\cite{pdg}.  

Our numerical analysis shows that the quantity $\Delta f_{\pi}$ is 
dominated by the second term in Eq. \ref{d2}, which comes from the 
four quark condensates. Thus, it is very sensitive to the vacuum dominance 
approximation adopted in our approach. 
The possible violation of this approximation was discussed in Ref. \cite{np}. 
Following Ref. \cite{np}, we introduce a factor $1+\kappa$ in front of the 
QED induced four quark condensates  based on the factorisation 
assumption in (\ref{r-}), (\ref{r0}), (\ref{p-}) and (\ref{p0}). 
A nonvanishing $\kappa$ signals the breakdown of the factorisation hypothesis. 
We find that $\Delta m_{\rho} $ is insensitive to $\kappa$ since it
 is dominated by the electromagnetic radiative 
correction. However, the $\Delta f_{\pi}$ is increased to $8.3$MeV with 
$\kappa=1$,  which is in good agreement with the PDG data. We estimate 
$\kappa =0.5 \sim 1.5$.  More accurate data on $\Delta f_{\pi}$ will 
determine how much the vacuum dominance approximation is being violated.

In summary,  we have presented a field theory calculation of the
electromagnetic effects in the isospin symmetry breakings of the
$\pi$ and $\rho$ mesons.  We have shown explicitly that the isospin symmetry
breaking induced by the elctromagnetic effects are dominant in the $\pi$ and
$\rho$ systems.   Our results show that the electromagnetic
effects in isospin symmetry breakings of pions are much larger than that
of $\rho$ mesons.  This is consistent with the available data, although
we have not included the effects due to the $\rho - \omega$ mixings
in the mass splittings of $\rho$ mesons.  Certainly, more accurate 
data are needed to test our theoretical approach, and to determine the
possible deviation from the factorization of the QED-induced 
four quark condensates.

This work is supported in part by the Natural Science Foundation 
and the Doctoral Program of State Education Commission of China.
The financial support by Peking University are also gratefully
 acknowledged. 

\newpage
\begin{center}
\section*{Appendix}
\end{center}
We present the expressions for the $\pi^0$ and $\rho^0$ systems in which 
the vacuum dominance approximation has not been implemented.
\begin{eqnarray}
\label{app1}
\Pi_{\rho^0} = 
- {1\over 4\pi^2} (1+{\alpha_s \over \pi} )\ln {-q^2 \over \mu^2}
- {1\over 4\pi^2} {e_u^2 + e_d^2 \over 2}{\alpha \over \pi} \ln {-q^2 \over \mu^2} \nonumber \\
+{1\over q^4} (m_d \langle {\bar d}d\rangle + m_u \langle {\bar u}u\rangle )
+{1\over 12q^4}\langle {\alpha_s \over \pi} G^2 \rangle \nonumber \\
+{1\over q^6}g_s^2 [ 
({\bar u}\gamma_\mu\gamma_5 {\lambda^a \over 2} u) 
({\bar u}\gamma^\mu\gamma_5 {\lambda^a \over 2} u)
+({\bar d}\gamma_\mu\gamma_5 {\lambda^a \over 2} d) 
({\bar d}\gamma^\mu\gamma_5 {\lambda^a \over 2} d) ] \nonumber \\
+{2\over 9q^6}g_s^2 [ 
({\bar u}\gamma_\mu {\lambda^a \over 2} u) 
( \sum\limits_{\mbox{q=u,d,s}}{\bar q}\gamma^\mu {\lambda^a \over 2} q)
+({\bar d}\gamma_\mu {\lambda^a \over 2} d) 
(\sum\limits_{\mbox{q=u,d,s}}{\bar q}\gamma^\mu {\lambda^a \over 2} q) ] \nonumber \\
+{1\over q^6}e^2 [ 
e_u^2({\bar u}\gamma_\mu\gamma_5  u) ({\bar u}\gamma^\mu\gamma_5  u)
+e_d^2 ({\bar d}\gamma_\mu\gamma_5 d) ({\bar d}\gamma^\mu\gamma_5 d) ] \nonumber \\
+{2\over 9q^6}e^2 [ 
(e_u{\bar u}\gamma_\mu  u) 
( \sum\limits_{\mbox{q=u,d,s}}e_q {\bar q}\gamma^\mu q)
+(e_d{\bar d}\gamma_\mu d) 
(\sum\limits_{\mbox{q=u,d,s}}e_q{\bar q}\gamma^\mu q) ] 
,
\end{eqnarray}
for the $\rho^0$ system and 
\begin{eqnarray}
\label{app2}
\Pi_{\pi^0}  = 
- {1\over 4\pi^2} (1+{\alpha_s \over \pi}) \ln {-q^2 \over \mu^2}
- {1\over 4\pi^2} {e_u^2 + e_d^2 \over 2}{\alpha \over \pi} \ln {-q^2 \over \mu^2} \nonumber \\
+{1\over q^4} (m_d \langle {\bar d}d\rangle + m_u \langle {\bar u}u\rangle )
+{1\over 12q^4}\langle {\alpha_s \over \pi} G^2 \rangle \nonumber \\
+{1\over q^6}g_s^2 [ 
({\bar u}\gamma_\mu {\lambda^a \over 2} u) 
({\bar u}\gamma^\mu {\lambda^a \over 2} u)
+({\bar d}\gamma_\mu {\lambda^a \over 2} d) 
({\bar d}\gamma^\mu {\lambda^a \over 2} d) ] \nonumber \\
+{2\over 9q^6}g_s^2 [ 
({\bar u}\gamma_\mu {\lambda^a \over 2} u) 
( \sum\limits_{\mbox{q=u,d,s}}{\bar q}\gamma^\mu {\lambda^a \over 2} q)
+({\bar d}\gamma_\mu {\lambda^a \over 2} d) 
(\sum\limits_{\mbox{q=u,d,s}}{\bar q}\gamma^\mu {\lambda^a \over 2} q) ] \nonumber \\
+{1\over q^6}e^2 [ 
e_u^2({\bar u}\gamma_\mu  u) ({\bar u}\gamma^\mu  u)
+e_d^2 ({\bar d}\gamma_\mu d) ({\bar d}\gamma^\mu d) ] \nonumber \\
+{2\over 9q^6}e^2 [ 
(e_u{\bar u}\gamma_\mu  u) 
( \sum\limits_{\mbox{q=u,d,s}}e_q {\bar q}\gamma^\mu q)
+(e_d{\bar d}\gamma_\mu  d) 
(\sum\limits_{\mbox{q=u,d,s}}e_q{\bar q}\gamma^\mu q) ] 
\end{eqnarray}
for the $\pi^0$ system. 
Similarly we may derive expressions for $\Pi_{\rho^{\pm}}$ 
and $\Pi_{\pi^{\pm}}$.

\vspace{5mm}
\newpage
\noindent {\bf Figure Captions}

\vspace{5mm}
\begin{itemize}
\begin{enumerate}
\item The two loop perturbative corrections for charge 
neutral currents
\item Additional diagrams generated by $J^e_{\mu}(x)$ for charged
quark systems. 
\item The electromagnetic interaction induced four quark condensates 
for charge neutral currents.
\item The additional four quark condensates for charged currents.
\item The Borel mass dependence of $\Delta m_{\rho}$.
\item The Borel mass dependence of $\Delta f_{\pi}$.
\end{enumerate}
\end{itemize}
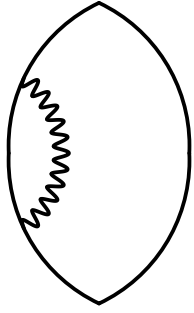
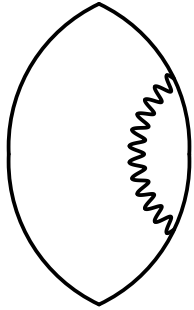
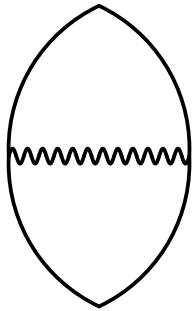
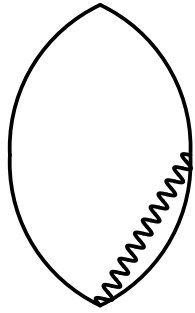
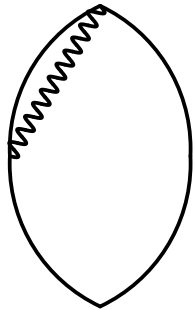
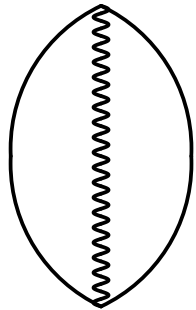
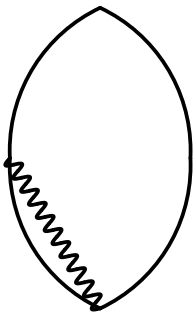
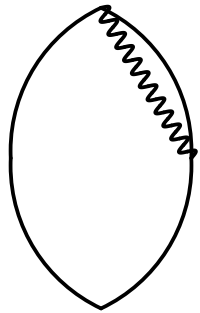
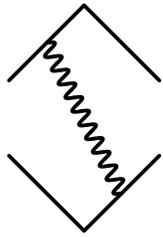
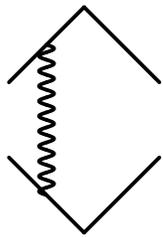
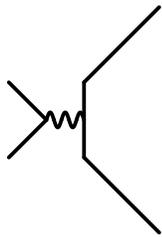
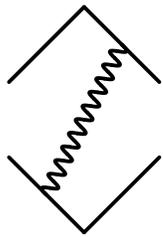
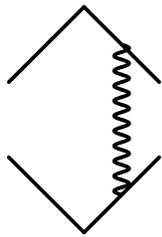
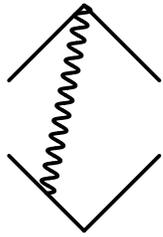
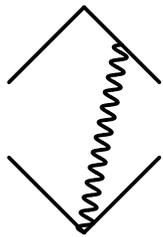
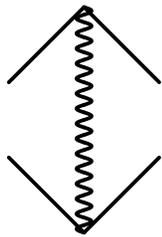
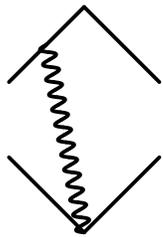
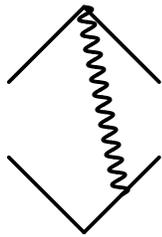
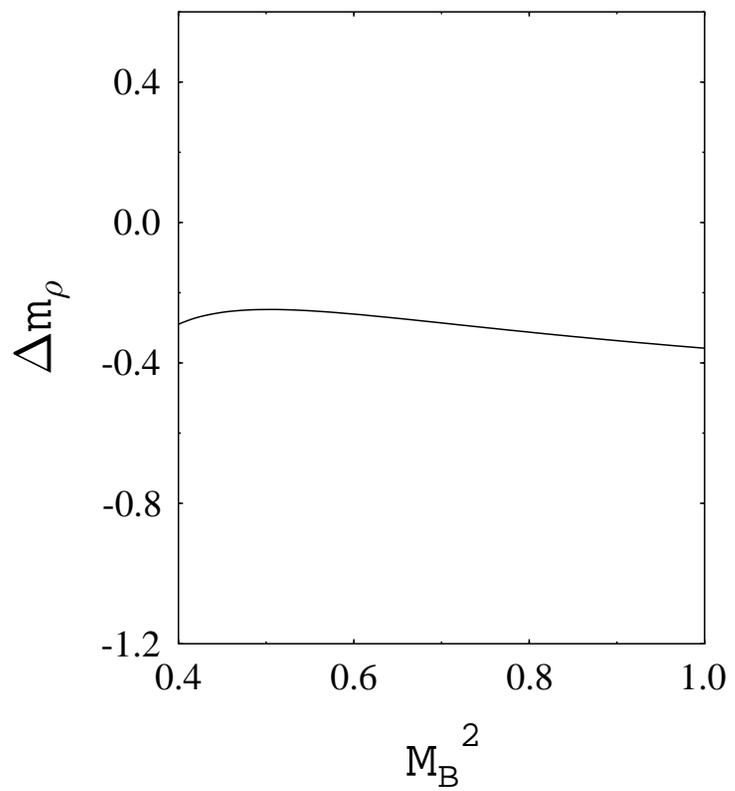
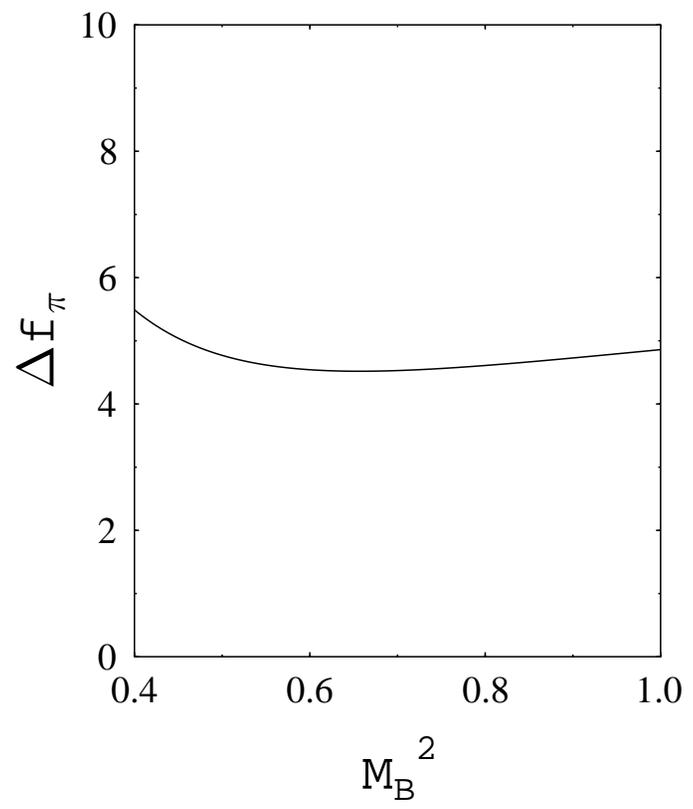
\end{document}